\documentclass[11pt]{article}% добавляется twoside для двусторонней печати

\usepackage[T1]{fontenc}
\usepackage[cp1251]{inputenc}
\usepackage{textcomp}
\usepackage[centertags]{amsmath}
\usepackage{amsfonts}
\usepackage{amssymb}
\usepackage[hypertex]{hyperref}
\usepackage{graphicx}
\usepackage{graphbox}
\usepackage[numbers,sort&compress]{natbib}% упорядочивает ссылки

\usepackage{paperinitial}% Установка параметров страницы

\paperinitialization{15mm}{15mm}{15mm}{15mm}{2pt}{10pt}

\DeclareMathOperator{\sgn}{sgn}

\DeclareMathOperator{\Texp}{Texp}

\newcommand{\lan}{\langle}
\newcommand{\ran}{\rangle}
\newcommand{\bs}{\boldsymbol}

\newcommand{\e}{\varepsilon}
\newcommand{\vf}{\varphi}

\newcommand{\s}{\sigma}

\newcommand{\al}{\alpha}
\newcommand{\be}{\beta}
\newcommand{\ga}{\gamma}

\newcommand{\de}{\delta}
\newcommand{\De}{\Delta}

\newcommand{\la}{\lambda}

\newcommand{\ups}{\upsilon}

\newcommand{\spx}{\mathbf{x}}
\newcommand{\spy}{\mathbf{y}}

\newcommand{\spp}{\mathbf{p}}
\newcommand{\spk}{\mathbf{k}}
\newcommand{\spe}{\mathbf{e}}

\begin{document}
\allowdisplaybreaks[4]% позволяет переносить многострочные формулы
\frenchspacing% уменьшение пробелов после запятых
%\setlength{\unitlength}{1pt}% устанавливает единицу длины в окружении picture
%\selectlanguage{english}

\title{{\Large\textbf{Transition radiation from a Dirac particle wave packet traversing a mirror}}}

\date{}

\author{P.O. Kazinski\thanks{E-mail: \texttt{kpo@phys.tsu.ru}}\;\, and G.Yu. Lazarenko\thanks{E-mail: \texttt{laz@phys.tsu.ru}}\\[0.5em]
{\normalsize Physics Faculty, Tomsk State University, Tomsk 634050, Russia}
}

\maketitle

\begin{abstract}

The explicit expression for the inclusive probability to record a photon created in transition radiation from a one Dirac particle wave packet traversing an ideally conducting plate is derived in the leading order of perturbation theory. The anomalous magnetic moment of the Dirac particle is taken into account. It is shown that the quantum corrections to transition radiation from an electrically charged particle give rise to production of photons with polarization vector orthogonal to the reaction plane ($E$-plane). These corrections result from both the quantum recoil and the finite size of a wave packet. As for transition radiation produced by a neutron falling normally onto the conducting plate, the probability to detect a photon with polarization vector lying in the reaction plane does not depend on the observation angle and the energy of the incident particle. The peculiarities of transition radiation stemming from different shapes of the particle wave packet are investigated. In particular, the transition radiation produced by the wave packet of one twisted Dirac particle is described. The comparison with classical approach to transition radiation is given and the quantum formula for the inclusive probability to detect a photon radiated by the $N$-particle wave packet is derived.

\end{abstract}

\section{Introduction}

Transition radiation from beams of charged particles is well studied both theoretically and experimentally \cite{GinzbThPhAstr,GaribYang,BazylZhev,Ginzburg,Pbook,RitEldI,FraAraBirII,Pafomov,BBCLS}. Nowadays this radiation is used in particle detectors, for diagnostics of particle beams, and as a simple source of photons with desired properties \cite{AndWess12,Pbook,BBCLS,Ginzburg,BKL5,BKL7}. The quantum theory of transition radiation was also studied \cite{GinzbThPhAstr,GaribYang,RitEldI}. However, the effects related to nontrivial form of the wave packets either were not taken into account or were considered in the framework of classical radiation theory with the classical currents produced by such wave packets \cite{IvKarl13prl,IvKarl13pra,KonPotPol,BliokhVErev}. In the present paper, we obtain the inclusive probability of transition radiation from the wave packet of one Dirac particle crossing an ideal conductor (a mirror) and investigate the dependence of this radiation on the form of the wave packet. The Dirac particle is assumed to be charged and have an anomalous magnetic moment. In particular, we derive the quantum formulas for the probability of transition radiation produced by the wave packets of electrons and neutrons including the wave packets with nonzero projection of the orbital angular momentum \cite{BliokhVErev,LBThY,Clark15,CapJachVin18,Sarena18,Sarena19}. It turns out that the polarization properties of this radiation differ from those given by the classical theory even in the case of vanishing orbital angular momentum. The transition radiation from a particle with anomalous magnetic moment also possesses rather uncommon features.

In order to simplify the calculations and to focus on the main peculiarities of transition radiation from wave packets, we consider the simplest model of transition radiation when the wave packet traverses an ideally conducting plate. This model is justified for good conductors in the infrared and optical spectrum ranges \cite{Ginzburg,Pbook,Pafomov}. The results of our study confirm the viewpoint existing in the literature \cite{MarcuseI,MarcuseII,SundMil90,LapFedEber93,PMHK08,CorsPeat11,WCCP16,PanGov18,Remez19} that the inclusive probability of radiation from a one-particle wave packet with nontrivial structure cannot be obtained with the aid of the classical approach based on the calculation of the classical field produced by the classical Dirac current in the wave zone. This is well seen by the example of transition radiation from twisted electrons that we study in detail in the present paper. Nevertheless, the probability of radiation calculated by the use of the classical formula proves to be physically meaningful. It describes the probability of the exclusive radiation process where the state of the escaping Dirac particle is also measured and it is measured in such a state as if it were evolving freely during the radiation process from the initial state of the Dirac particle. The measurement of the Dirac particle in that state suppresses completely the quantum recoil due to photon radiation. Furthermore, the classical formula gives the correct result for the inclusive probability of coherent radiation from a beam of particles in the leading order of perturbation theory \cite{MarcuseII}. We deduce the general formula for the inclusive probability of radiation from $N$-particle wave packet of fermions (for $N=2$ see, e.g., \cite{AngDiPiaz18,AngDiPiaz19}) and point out the conditions when the classical formula for radiation probability is reproduced.

Other results of the present paper are discussed in detail in Conclusion. Therefore, we will not dwell on it here. The paper is organized as follows. In Sec. \ref{GenForm}, we collect the general formulas regarding quantum electrodynamics (QED) of Dirac particles with anomalous magnetic moment in the Coulomb gauge in the presence of the ideally conducting plate. Section \ref{RadWavPack} is devoted to derivation of the general formula for inclusive probability of transition radiation from the wave packet of one Dirac particle. In Sec. \ref{NormFall}, we investigate the particular case of this general formula for a normal incidence of the one-particle wave packet. In Sec. \ref{WavePack}, the wave packets of various profiles are considered. In particular, the transition radiation from twisted wave packets of electrons and neutrons is scrutinized in Sec. \ref{TwisPart}. A comparison of the quantum and classical approaches to radiation from wave packets is given in Appendix \ref{ClassAppr}. Appendix \ref{UsefForm} contains some ponderous formulas used in the main text.

We use the system of units such that $\hbar=c=1$ and $e^2=4\pi\al$, where $\al$ is the fine structure constant. The Minkowski metric is taken with the mostly minus signature.

%\newpage
\section{General formulas}\label{GenForm}

The Lagrangian density of the system at issue reads
\begin{equation}\label{action_model}
    \mathcal{L}=-\frac{1}{4}F_{\mu\nu}F^{\mu\nu}+\bar{\psi}[\ga^\mu(i\partial_\mu-eA_\mu)-m]\psi -\frac{\mu_a}{2}F_{\mu\nu}\bar{\psi}\s^{\mu\nu}\psi,
\end{equation}
where $\mu_a=ae/(2m)=:\sgn(e)a\mu_B$ is the anomalous magnetic moment, $e$ is the particle charge, $m$ is the particle mass, $F_{\mu\nu}:=\partial_{[\mu}A_{\nu]}$ is the strength tensor of the electromagnetic field, and $\psi$ is the Dirac spinor. We use the notation and conventions adopted in \cite{BjoDre,Okun,BaKaStrbook} and
\begin{equation}
    \s^{\mu\nu}:=\frac{i}{2}[\ga^\mu,\ga^\nu],\qquad\ga^5=-i\ga^0\ga^1\ga^2\ga^3.
\end{equation}
Fixing the Coulomb gauge and excluding the field $A_0$, we come to
\begin{equation}\label{action_model1}
    \mathcal{L}_{\text{Coul}}=\frac12\dot{A}^2_i-\frac14F_{ij}F_{ij}+\bar{\psi}(i\ga^\mu\partial_\mu-m)\psi -eA_i\bar{\psi}\ga^i\psi -\mu\dot{A}_i\bar{\psi}\s^{0i}\psi-\frac{\mu_a}{2}F_{ij}\bar{\psi}\s^{ij}\psi-V_{\text{Coul}},
\end{equation}
where
\begin{equation}
    V_{\text{Coul}}=-\frac12[e\bar{\psi}\ga^0\psi+\mu_a\partial_i(\bar{\psi}\s^{0i}\psi)]\De^{-1}[e\bar{\psi}\ga^0\psi+\mu_a\partial_i(\bar{\psi}\s^{0i}\psi)],
\end{equation}
and $\De^{-1}$ is the inverse to Laplace operator. The model \eqref{action_model}, \eqref{action_model1} can be employed for description of the electromagnetic radiation created by electrons, protons, neutrons, and neutrinos. In the case of neutrinos, one has to take into account the three flavors of neutrinos, their mixing, and their interaction with electrons of the medium \cite{LobStud03,IonIonKaz11,DOlivLoz12,Studenik16}.

We assume that the ideally conducting plate is located at $z\leqslant0$ and the detector of photons is placed in the region $z>0$ sufficiently far from the plate. Then the quantum fields in the interaction picture are written as
\begin{equation}
\begin{split}
    \hat{\mathbf{A}}(x)&=\sum_\la \int_+\frac{Vd\spk}{(2\pi)^3}\frac{1}{\sqrt{2k_0V}}\big[\mathbf{a}_\la(k_3;z)e^{-ik_0x^0+i\spk_\perp \spx_\perp}\hat{c}_\la(\spk) +\text{H.c.}\big],\\
    \hat{\psi}(x)&=\sum_s\int\frac{Vd \spp}{(2\pi)^3}\sqrt{\frac{m}{Vp_0}}\big[u_s(\spp)e^{-ip_\mu x^\mu}\hat{a}_s(\spp) +\ups_s(\spp)e^{ip_\mu x^\mu}\hat{b}^\dag_s(\spp)\big],\\
    \hat{\bar{\psi}}(x)&=\sum_s\int\frac{Vd \spp}{(2\pi)^3}\sqrt{\frac{m}{Vp_0}}\big[\bar{u}_s(\spp)e^{ip_\mu x^\mu} \hat{a}^\dag_s(\spp) +\bar{\ups}_s(\spp)e^{-ip_\mu x^\mu}\hat{b}_s(\spp)\big],\\
\end{split}
\end{equation}
where $k_0=|\spk|$, $p_0=\sqrt{m^2+\spp^2}$, $V$ characterizes the normalization volume, the plus sign in the integral limits means that the integration region is restricted by the inequality $k^3\geqslant0$, and the indices $s$ and $\la$ run the two values. The creation-annihilation operators obey the standard (anti)commutation relations:
\begin{equation}
\begin{gathered}\relax
    [\hat{c}_\la(\spk),\hat{c}^\dag_{\la'}(\spk')]=\de_{\ga,\ga'}\equiv\frac{(2\pi)^3}{V}\de_{\la,\la'}\de(\spk-\spk'),\\ [\hat{a}_s(\spp),\hat{a}^\dag_{s'}(\spp')]=\de_{\al,\al'}\equiv\frac{(2\pi)^3}{V}\de_{s,s'}\de(\spp-\spp'),\qquad [\hat{b}_s(\spp),\hat{b}^\dag_{s'}(\spp')]=\de_{\al,\al'}\equiv\frac{(2\pi)^3}{V}\de_{s,s'}\de(\spp-\spp'),
\end{gathered}
\end{equation}
where the square brackets denote a graded commutator. Other graded commutators of the creation-annihilation operators vanish. Henceforth, for brevity we use the notation
\begin{equation}
\begin{gathered}
    \al=(s,\spp),\qquad \al'=(s',\spp'),\qquad \ga:=(\la,\spk), \\
    \sum_\al\equiv\sum_s\int\frac{Vd \spp}{(2\pi)^3},\qquad \sum_{\al'}\equiv\sum_{s'}\int\frac{Vd \spp'}{(2\pi)^3},\qquad \sum_\ga\equiv\sum_\la \int_+\frac{Vd\spk}{(2\pi)^3},
\end{gathered}
\end{equation}
and so on.

The mode functions of the electromagnetic field take the form
\begin{equation}
\begin{aligned}
    \mathbf{a}_\la(k_3;z)&=\sum_{r=\pm1}\mathbf{f}^{(\la)}_r(\spk) e^{irk^3z},\qquad \mathbf{f}^{(\la)}_r(\spk)=r\mathbf{f}^{(\la)}(\spk)+(1-r)\spe_3f^{3(\la)}(\spk),&\quad&\text{for $z>0$};\\
    \mathbf{a}_\la(k_3;z)&=0,&\quad&\text{for $z<0$};
\end{aligned}
\end{equation}
where $\spe_3$ is the unit vector along the $z$ axis. The polarization vector of a photon with helicity $\la$ is written as
\begin{equation}
    \mathbf{f}^{(\la)}(\spk)=(\cos\phi\cos\theta-i\la\sin\phi,\sin\phi\cos\theta+i\la\cos\phi,-\sin\theta)/\sqrt{2},
\end{equation}
where $k_\perp=|k^1+ik^2|$, $\phi=\arg (k^1+ik^2)$, $\sin\theta:=k_\perp/k_0\equiv n_\perp$, $\cos\theta:=k^3/k_0\equiv n^3$:
\begin{equation}
    \spk=k_0\mathbf{n}=k_0(\sin\theta\cos\phi,\sin\theta\sin\phi,\cos\theta).
\end{equation}
The mode functions of the Dirac spinors are defined as in \cite{BjoDre,Okun,BaKaStrbook}:
\begin{equation}\label{mode_fun_Dir}
    u_s(\spp)=\frac{m+\hat{p}}{\sqrt{2m(p_0+m)}}
    \left[
      \begin{array}{c}
        \chi_s \\
        0 \\
      \end{array}
    \right],\qquad
    \ups_s(\spp)=\frac{m-\hat{p}}{\sqrt{2m(p_0+m)}}
    \left[
      \begin{array}{c}
        0 \\
        \chi_s \\
      \end{array}
    \right],
\end{equation}
where $\hat{p}:=\ga^\mu p_\mu$. In particular,
\begin{equation}
\begin{aligned}
    \sum_{s}u_s(\spp)\bar{u}_s(\spp)&=\frac{m+\hat{p}}{2m},&\qquad \sum_{s}\ups_s(\spp)\bar{\ups}_s(\spp)&= -\frac{m-\hat{p}}{2m},\\
    u_s(\spp)\bar{u}_s(\spp)&=\frac{m+\hat{p}}{2m}\frac{1-\ga^5\hat{s}}{2},&\qquad \ups_s(\spp)\bar{\ups}_s(\spp)&=-\frac{m-\hat{p}}{2m}\frac{1-\ga^5\hat{s}}{2},
\end{aligned}
\end{equation}
where the spin vector
\begin{equation}\label{smu_zeta}
    s^\mu=\Big(\frac{\bs{\zeta}\spp}{m},\bs{\zeta}+\frac{\spp(\bs{\zeta}\spp)}{m(p_0+m)}\Big),\qquad s^\mu p_\mu=0,\qquad s^2=-\bs{\zeta}^2,
\end{equation}
and the vector $\bs{\zeta}$ specifies the spin of a Dirac particle in the rest frame. For pure states $\bs{\zeta}^2=1$.

The evolution operator is expressed through the $\hat{S}$-operator as
\begin{equation}\label{U_oper}
    \hat{U}_{t_2,t_1}=\hat{U}^0_{t_2,0}\hat{S}_{t_2,t_1} \hat{U}^0_{0,t_1},
\end{equation}
where the operator $\hat{U}^0_{t_2,t_1}$ describes a free evolution and
\begin{equation}\label{S_oper}
    \hat{S}_{t_2,t_1}=\Texp\Big\{-i\int_{t_1}^{t_2}dx:[e\hat{A}_i\hat{\bar{\psi}}\ga^i\hat{\psi} +\mu_a\dot{\hat{A}}_i\hat{\bar{\psi}}\s^{0i}\hat{\psi} +\frac{\mu_a}{2} \hat{F}_{ij}\hat{\bar{\psi}}\s^{ij}\hat{\psi}](x): -i\int_{t_1}^{t_2}dt\hat{V}_{\text{Coul}}(t) \Big\}.
\end{equation}
All the operators in the $T$-exponent are taken in the interaction picture. The Coulomb term in the interaction Hamiltonian becomes
\begin{equation}
    \hat{V}_{\text{Coul}}(t)=-\frac12\int d\spx d\spy:[e\hat{\bar{\psi}}\ga^0\hat{\psi}+\mu_a\partial_i(\hat{\bar{\psi}}\s^{0i}\hat{\psi})](t,\spx): \De^{-1}(\spx-\spy) :[e\hat{\bar{\psi}}\ga^0\hat{\psi}+\mu_a\partial_i(\hat{\bar{\psi}}\s^{0i}\hat{\psi})](t,\spy):,
\end{equation}
where $\De^{-1}(\spx-\spy)$ is the kernel of the operator $\De^{-1}$. The contribution of the Coulomb interaction to the process we are investigating is of a higher order of smallness with respect to the coupling constants than the leading contribution of the term in the square brackets in the $T$-exponent in \eqref{S_oper}.

\section{Radiation from a wave packet}\label{RadWavPack}

The state of the Dirac particle at the instant of time $t_1$ has the form
\begin{equation}\label{ini_state}
    |\vf\ran:=\sum_\al\sqrt{\frac{(2\pi)^3}{V}} \tilde{\vf}_\al|\al\ran,\qquad \sum_\al\frac{(2\pi)^3}{V} |\tilde{\vf}_\al|^2=\sum_s\int d\spp|\tilde{\vf}_s(\spp)|^2=1.
\end{equation}
As follows from the problem statement, $\tilde{\vf}_s(\spp)$ is concentrated at $p^3<0$, i.e., the wave packet of a particle falls onto the conducting plate situated at $z=0$. Here we assume that, for a given momentum of the incident particle, one can neglect its diffraction on the atoms of the mirror plate. Furthermore, the model of a conducting plate used by us implies that bremsstrahlung can be neglected in the region of energies where the photons are detected. As far as electrons are concerned, these conditions are fulfilled when the electron kinetic energies are larger or of order $50$ keV \cite{BBCLS,Pafomov,FraAraBirII,Pbook}. As for protons and neutrons, the applicability conditions are satisfied when the de Broglie wavelengths of these particles are small in comparison with the interatomic distances in the mirror plate. Besides, the diffraction and bremsstrahlung can be completely suppressed if one perforates the conducting plate making the channel for the incident particles. The transverse width of the channel, $d_\perp$, must satisfy
\begin{equation}
    k_\perp d_\perp\ll1,
\end{equation}
and, of course, it should be larger than the transverse size of the particle wave packet.

The leading contribution to the amplitude of the process of photon radiation by a Dirac particle is given by the matrix element
\begin{equation}\label{matrix_elem}
    A(\ga,\al';\al):=\lan \spk ,\la;\spp',s'|\hat{U}_{t_2,t_1}|\spp,s\ran.
\end{equation}
Substituting \eqref{U_oper} into this expression, retaining only the terms that are of leading order with respect to the coupling constants, and using the relations
\begin{equation}
    \hat{U}^0_{0,t}\hat{a}_\al \hat{U}^0_{t,0}=e^{-iE_\al t}\hat{a}_\al,\qquad \hat{U}^0_{t_2,t_1}|0\ran=e^{-iE_0(t_2-t_1)}|0\ran,
\end{equation}
where $E_\al$ is the energy of the one-particle state $\al$ and $E_0$ is the vacuum energy, the transition amplitude can be cast into the form
\begin{equation}
\begin{split}
    A(\ga,\al';\al)=&\,-i\frac{m}{V}e^{-iE_0(t_2-t_1)}e^{-i(k_0+p_0')t_2+ip_0t_1} \times\\
    &\times\int_{t_1}^{t_2}dx a^*_{i\la}(x_3)\bar{u}_{\al'}\big[e\ga^i -i\mu_a(k_0\s^{i0} +(p_j-p'_j)\s^{ij}) \big] u_\al \frac{e^{ik_0x^0-i\spk_\perp \spx_\perp+i(p_\mu'-p_\mu)x^\mu}}{\sqrt{2Vk_0p_0p_0'}},
\end{split}
\end{equation}
where the integration by parts was carried out in the contribution of the anomalous magnetic moment.

We are interested in the transition amplitude from the state \eqref{ini_state}. It is written as
\begin{equation}
    A(\ga,\al';\vf):=\sum_\al \sqrt{\frac{(2\pi)^3}{V}}\tilde{\vf}_\al A(\ga,\al';\al).
\end{equation}
Denote as
\begin{equation}\label{free_evol}
    \vf_\al:=e^{ip_0t_1}\tilde{\vf}_\al.
\end{equation}
The function $\vf_\al$ determines the form of the Dirac particle wave packet at the instant of time $t=0$ in the leading order of perturbation theory, i.e., according to \eqref{free_evol}, the particle wave packet evolves freely from the moment $t=t_1$ to the moment $t=0$. From the experimentalists point of view, it is convenient to specify the form of the particle wave packet at the instant of time $t=0$. Therefore, henceforward we suppose that the function $\vf_\al$ is given and $\tilde{\vf}_\al$ is found from \eqref{free_evol}. Obviously, $\vf_\al$ also satisfies the normalization condition \eqref{ini_state}.

Then
\begin{equation}\label{ampl_phi}
\begin{split}
    A(\ga,\al';\vf)=&\,-i\frac{m}{V}e^{-iE_0(t_2-t_1)}e^{-i(k_0+p_0')t_2}\sum_\al \frac{(2\pi)^{3/2}}{V^{1/2}} \times\\
    &\times\int_{t_1}^{t_2}dx a^*_{i\la}(x_3)\bar{u}_{\al'}\big[e\ga^i -i\mu_a(k_0\s^{i0} +(p_j-p'_j)\s^{ij}) \big] u_\al \vf_\al \frac{e^{ik_0x^0-i\spk_\perp \spx_\perp+i(p_\mu'-p_\mu)x^\mu}}{\sqrt{2Vk_0p_0p_0'}}.
\end{split}
\end{equation}
On squaring the modulus of \eqref{ampl_phi}, the phase factors standing before the sum sign in \eqref{ampl_phi} disappear. Up to these factors, putting $t_1\rightarrow-\infty$ and $t_2\rightarrow+\infty$, we come to
\begin{equation}\label{ampl_phi1}
\begin{split}
    A(\ga,\al';\vf)=\,&m\frac{(2\pi)^{3/2}}{V^{1/2}}\sum_{s}\int d\spp\de(k_0+p_0'-p_0)\de(\spk_\perp+\spp'_\perp-\spp_\perp)\times\\
    &\times
    \int dx_3 \frac{e^{i(p_3-p'_3)x_3}}{\sqrt{2Vk_0p_0p_0'}} a^*_{i\la}(x_3)\bar{u}_{\al'}\big[e\ga^i -i\mu_a(k_0\s^{i0} +(p_j-p'_j)\s^{ij}) \big] u_\al\vf_\al.
\end{split}
\end{equation}
The integral over $x^3$ is readily evaluated
\begin{equation}\label{int_x3}
    \int_{-\infty}^\infty dx^3 e^{i(p_3-p'_3)x_3}\mathbf{a}^*_\la(x_3)=\int_{0}^\infty dx^3 e^{i(p_3-p'_3)x_3}\mathbf{a}^*_\la(x_3)=\sum_{r=\pm1}\frac{-i\mathbf{f}^{(\la)}_r(\spk)}{p_3'-p_3+rk_3-i0}.
\end{equation}
As long as the conducting plate violates the translation invariance along the $z$ axis, the conservation law of the momentum projection onto this axis does not hold.

The inclusive probability to record a photon in the leading order of perturbation theory reads
\begin{equation}\label{dP_ini}
    dP(\la,\spk):=\sum_{\al'} |A(\ga,\al';\vf)|^2\frac{Vd\spk}{(2\pi)^3}.
\end{equation}
The comparison of this formula with the classical approach for description of radiation from a one-particle wave packet is presented in Appendix \ref{ClassAppr}. Apart from the sum over the final states of a Dirac particle, the expression \eqref{dP_ini} contains the two sums over the quantum numbers of the initial state
\begin{equation}
    \sum_{\al,\bar{\al}},\qquad\al=(s,\spp),\;\bar{\al}=(\bar{s},\bar{\spp}).
\end{equation}
The delta functions entering into \eqref{ampl_phi1} allow one to remove six of the nine integrations in \eqref{dP_ini}. Then accounting for the delta functions
\begin{multline}
    \de(k_0+p_0'-p_0)\de(\spk_\perp+\spp'_\perp-\spp_\perp) \de(k_0+p_0'-\bar{p}_0)\de(\spk_\perp+\spp'_\perp-\bar{\spp}_\perp)= \de(p_0-\bar{p}_0)\de(\spp_\perp-\bar{\spp}_\perp)\times\\
    \times \de(k_0+p_0'-p_0)\de(\spk_\perp+\spp'_\perp-\spp_\perp),
\end{multline}
we come to
\begin{equation}\label{p'_mom}
\begin{split}
    p_0'&=p_0-k_0,\\
    \spp'_\perp&=\spp_\perp-\spk_\perp,\\
    p'^3&=\s\sqrt{(p_0-k_0)^2-(\spp_\perp-\spk_\perp)^2-m^2}=\s\sqrt{p_3^2-2k_0p_0+2\spk_\perp \spp_\perp +k_3^2},\quad\s=\pm1,
\end{split}
\end{equation}
and
\begin{equation}
    \bar{p}^3=\pm\sqrt{p_0^2-\spp_\perp^2-m^2},\qquad \bar{p}^0=p^0,\qquad\bar{\spp}_\perp=\spp_\perp.
\end{equation}
The contribution of the root $\bar{p}^3=\sqrt{p_0^2-\spp_\perp^2-m^2}$ to \eqref{dP_ini} is negligibly small since $\vf_{\bar{s}}(\bar{\spp})$ is concentrated at negative $\bar{p}^3$. Therefore, the inclusive probability \eqref{dP_ini} becomes
\begin{equation}\label{dP_ini1}
\begin{split}
    dP(\la,\spk;\vf)=&\,m\sum_{r,r',\s,s,\bar{s}}\int\frac{d\spp}{|p_3'p_3|} \frac{\vf_s(\spp)\vf^*_{\bar{s}}(\spp) f^{(\la)}_{ri}(\spk) f^{*(\la)}_{r'j}(\spk)}{2(p_3'-p_3+rk_3-i0)(p_3'-p_3+r'k_3+i0)}\times\\
    &\times\bar{u}_{\bar{s}}(\spp)\big[e\ga^i +i\mu_a(p'_\nu-p_\nu)\s^{\nu i}\big](m+\hat{p}')\big[e\ga^j -i\mu_a(p'_\rho-p_\rho)\s^{\rho j}\big]u_{s}(\spp) \frac{d\spk}{2(2\pi)^3k_0}.
\end{split}
\end{equation}
The contribution of the root with $\s=1$ corresponds to the scattering of the Dirac particle on its image in the mirror. It is not difficult to show using the mass-shell condition that $p_3'-p_3+rk_3<0$. Hence, the imaginary additions $\pm i0$ can be omitted in the denominators.

Let us find the estimates when the approximations used above such as the approximation of an infinite conducting plate, proceeding to the limit $|t_{1,2}|\rightarrow\infty$, and the use of the model of a sharp boundary of a mirror are valid. Hereinafter we assume that the quantum recoil is small, viz.,
\begin{equation}\label{quant_rec}
    k_0/(p_3\be_3)\ll1,
\end{equation}
where $\be_3=p_3/(m\ga)$ and $\ga$ is the Lorentz factor. Then
\begin{equation}
    p_3'\approx-\s[p_3-k_0(1-\bs{\be}_\perp\mathbf{n}_\perp)/\be_3].
\end{equation}
Let $p^\vf_3$, $p_\perp^\vf$ be the minimum typical scales of variations of $\vf_s(p_3,\spp_\perp)$, $L_\perp$ be the typical transverse size of the plate, and $T=t_2-t_1$. The integral over $x^3$ in \eqref{int_x3} is saturated on the scale
\begin{equation}
    \ell_z=
    \left\{
      \begin{array}{ll}
        \dfrac{|\be_3|}{k_0(1-\bs{\be}_\perp\mathbf{n}_\perp+rn^3)}, & \hbox{$\s=-1$;} \\
        1/(2p_3), & \hbox{$\s=1$,}
      \end{array}
    \right.
\end{equation}
where $\bs{\be}_\perp$ are the components of particle velocity along the conducting plate, $\mathbf{n}_\perp:=\spk_\perp/k_0$, and $n^3=k^3/k_0$. Then, in terms of the introduced quantities, the model employed is applicable when the diffraction of incident particles and the created bremsstrahlung can be neglected, as we discussed above, and the following estimates are met:
\begin{equation}
    T|\be_3|\gg \ell_z,\qquad T |\be_3| p_3^\vf\gtrsim 1,\qquad L_\perp p_\perp^\vf\gtrsim1.
\end{equation}
The vacuum-mirror boundary can be considered as sharp provided the thickness of the transition layer, $\de$, is much less than  $\ell_z$. In the opposite case, $\de\gg\ell_z$, the corresponding contribution to the integral defining the amplitude is exponentially suppressed by the factor of the form $\exp(-a\de/\ell_z)$, where $a$ is some positive constant of order unity. In what follows we assume that $\de\gg\ell_z$ for $\s=1$ and $\de\ll\ell_z$ for $\s=-1$. This situation is realized for electrons with kinetic energy larger than $1$ keV and for protons and neutrons with kinetic energy larger than $1$ eV. In that case, one can neglect the contribution with $\s=1$ in \eqref{dP_ini1}. From the physical viewpoint, it means that we neglect the contribution to transition radiation from the Dirac particles reflected from the mirror due to scattering on their image in it. Therefore, henceforth $p'^3$ is defined by formula \eqref{p'_mom} with $\s=-1$.

The expression \eqref{dP_ini1} for the radiation probability contains the positive definite Hermitian density matrix
\begin{equation}
    R_{s,\bar{s}}(\spp,\bar{\spp}):=\vf_s(\spp)\vf^*_{\bar{s}}(\bar{\spp}).
\end{equation}
Define the spin density matrix as
\begin{equation}
    \rho_{s,\bar{s}}(\spp,\bar{\spp}):=R_{s,\bar{s}}(\spp,\bar{\spp})/c(\spp),\qquad c(\spp):=\sum_sR_{s,s}(\spp,\spp),\qquad \int d\spp c(\spp)=1.
\end{equation}
Formula \eqref{dP_ini1} is generalized to an arbitrary mixed initial state of the Dirac particle by the replacement
\begin{equation}\label{dens_matr}
    \vf_s(\spp)\vf^*_{\bar{s}}(\spp)\rightarrow R_{s,\bar{s}}(\spp,\spp)=c(\spp) \rho_{s,\bar{s}}(\spp,\spp),
\end{equation}
where $R_{s,\bar{s}}(\spp,\spp)$ is the diagonal in the momentum space of the density matrix of the initial mixed state. One can show that (see, e.g., \cite{BaKaStrbook,AkhBerQED} and Sec. \ref{SpinPolSt} below)
\begin{equation}\label{mix_state}
    \sum_{s,\bar{s}}\rho_{s,\bar{s}}(\spp,\spp)u_{s}(\spp)\bar{u}_{\bar{s}}(\spp)=\frac{m+\hat{p}}{2m}\frac{1-\ga^5\hat{s}}{2},
\end{equation}
where the effective spin $4$-vector $s^\mu$ has the form \eqref{smu_zeta} with a certain vector $\bs{\zeta}(\spp)$ such that $|\bs{\zeta}|\leqslant1$. In the case of a pure state $|\bs{\zeta}|=1$, whereas for a mixed state $\bs{\zeta}^2<1$. In particular, for the mixed states corresponding to a natural spin polarization, where the opposite values of the spin are equally probable, $\bs{\zeta}=0$ and, consequently, $s_\rho=0$. If $\rho_{s,\bar{s}}(\spp,\spp)$ does not depend on $\spp$, then $\bs{\zeta}$ is also independent of $\spp$. The concrete examples of the states are considered below in Sec. \ref{WavePack}.

Notice that, in virtue of the symmetries of the problem we are studying and the approximations made, the expression for the probability of radiation of a photon \eqref{dP_ini1} depends only on the momentum space diagonal of the density matrix of the incident particle. This, in particular, shows that \eqref{dP_ini1} is independent of the common phase factor of the wave functions $\vf_s(\spp)$ constituting the density matrix, i.e., the transform,
\begin{equation}
    \vf_s(\spp)\rightarrow \vf_s(\spp) e^{i\xi(\spp)}, \qquad\forall\xi(\spp)\in \mathbb{R},
\end{equation}
is a symmetry of \eqref{dP_ini1}.

Substituting \eqref{dens_matr}, \eqref{mix_state} into \eqref{dP_ini1} and evaluating the trace of the respective products of the $\ga$-matrices, we obtain
\begin{equation}\label{dP_fin}
\begin{split}
    dP(\la,\spk;\vf)=&\,\frac{d\spk}{32\pi^3k_0}\int\frac{d\spp c(\spp)}{|p_3'p_3|} \sum_{r,r'}\frac{ f^{(\la)}_{ri}(\spk)f^{*(\la)}_{r'j}(\spk)}{(p_3'-p_3+rk_3)(p_3'-p_3+r'k_3)}\times\\
    &\times\Big\{e^2\big(p^{(i}p'^{j)} +\eta^{ij}q^2/2 +imq_\mu s_\nu\e^{\mu\nu ij} \big)+\\
    &+e\mu_a\big(2mq^2\eta^{ij} -2mq^iq^j +i[2m^2q_\mu s_\nu+(qs)p_\mu q_\nu+q^2q_\mu s_\nu/2]\e^{\mu\nu ij}\big)+\\ &+\mu_a^2\big(q^2[2m^2\eta^{ij}-(p^i+p'^i)(p^j+p'^j)/2]-2m^2q^iq^j +2im[(qs)p_\mu q_\nu +q^2q_\mu s_\nu/2]\e^{\mu\nu ij}\big) \Big\},
\end{split}
\end{equation}
where $\e^{0123}=1$, $\eta^{ij}=-\de_{ij}$, and
\begin{equation}
\begin{gathered}
    q_\mu:=p_\mu-p'_\mu,\qquad q^2=2(m^2-pp')=k_3^2-(p_3-p_3')^2<0,\\
    q^\mu=k^\mu_r+\de^\mu_3(q^3-rk^3).
\end{gathered}
\end{equation}
The vector $k^\mu_r$ possesses the same components as $k^\mu$ but with $k^3\rightarrow rk^3$. It is clear that the expression in the curly brackets in \eqref{dP_fin} contracted with the polarization vectors can be written in an explicitly Lorentz-invariant form. The terms depending on the effective particle spin contribute only to a circular polarization of radiated photons. They vanish when the polarization vector $\mathbf{f}^{(\la)}(\spk)$ is real and so corresponds to a certain linear polarization. The explicit expression for the sum over $r$, $r'$ on the first line in \eqref{dP_fin} is given in formula \eqref{contractions}. The classical transition radiation from a point charged particle without magnetic moment is described by the first term in the curly brackets standing at $e^2$ in \eqref{dP_fin} provided one replaces $p'^{i}$ by $p^{i}$ in it and employs the small quantum recoil approximation (see Appendix \ref{UsefForm}) in the factor at the curly brackets.

\section{Normal falling}\label{NormFall}

Let us consider a particular case of the general formula \eqref{dP_fin} when the wave packet crosses normally the surface of the conducting plate. In this section, we shall find the probability to detect a photon with linear polarization along the vector $[\spp,\spk]$ in transition radiation from the wave packet of one Dirac particle. Besides, we shall derive the probability to record a photon in this radiation when its polarization is not measured.

As is well known (see, e.g., \cite{Ginzburg}), the transition radiation from a classical point particle with charge $e$ is linearly polarized in the plane spanned by the vectors $\spp$ and $\spk$. However, it is not difficult to see that this property does not hold for the quantum expression \eqref{dP_fin} even in the case $s_\rho=0$ and $\mu_a=0$. For simplicity, we suppose that $c(\spp)$ is concentrated near the set of points
\begin{equation}\label{norm_fall}
    \spp=(0,0,p),\qquad p<0,
\end{equation}
with different $p$. In other words, we will consider a normal incidence of the wave packet on the mirror. The concrete form of the wave packets meeting such requirements will be given in Sec. \ref{WavePack}.

Consider the probability to detect a photon with the polarization vector
\begin{equation}
    f_3=0,\qquad\mathbf{f}\in\mathbb{R}.
\end{equation}
In that case, $\mathbf{f}$ is proportional to $[\spe_3,\spk]$. Assume for a while that the condition \eqref{norm_fall} is not fulfilled. Then
\begin{equation}
    \mathbf{f}_r=r\mathbf{f},
\end{equation}
the terms containing $s_\rho$ vanish, and the other terms in \eqref{dP_fin} are simplified with the aid of formulas \eqref{contractions}. As a result, we come to
\begin{equation}\label{dP_phi}
    dP(\la,\spk;\vf)=\int\frac{d\spp c(\spp)}{|p_3'p_3|(q^2)^2}
    \big[4(e^2-q^2\mu_a^2)(\spp\mathbf{f})^2-q^2(e+2m\mu_a)^2\big]\frac{k_3^2d\spk}{16\pi^3k_0}.
\end{equation}
Now we suppose that $c(\spp)$ is concentrated near the points \eqref{norm_fall}. In that case,
\begin{equation}\label{dP_phi1}
    dP(\la,\spk;\vf)=-(e+2m\mu_a)^2\int\frac{d\spp c(\spp)}{|p_3'p_3|q^2}\frac{k_3^2d\spk}{16\pi^3k_0}.
\end{equation}
Notice that $e+2m\mu_a=e(1+a)$ for an electrically charged particle. The expression obtained is valid for any effective spin vector $s_\rho$ including $s_\rho=0$.

Neglecting the higher orders in the quantum recoil parameter \eqref{quant_rec} and using the formulas \eqref{small_reco}, we deduce
\begin{equation}\label{dP_trans0}
    dP(\la,\spk;\vf)=\frac{(e+2m\mu_a)^2}{m^2} \int\frac{d\spp c(\spp)}{n_3^2+n_\perp^2\ga^2(\spp)}  \frac{n_3^2d\spk}{16\pi^3k_0}.
\end{equation}
In the nonrelativistic limit, $\ga^2\approx1$, this expression is written as
\begin{equation}
    dP(\la,\spk;\vf)=\frac{(e+2m\mu_a)^2}{m^2} \frac{n_3^2d\spk}{16\pi^3k_0}.
\end{equation}
It does not depend on the profile of the wave packet along $p_3$. In the general case, if $c(\spp)$ possesses only one sharp peak near some point of the form \eqref{norm_fall}, we approximately have
\begin{equation}\label{dP_trans}
    dP(\la,\spk;\vf)\approx\frac{(e+2m\mu_a)^2}{m^2(n_3^2+n_\perp^2\ga^2)} \frac{n_3^2d\spk}{16\pi^3k_0}.
\end{equation}
Notice that \eqref{dP_phi1} is not zero for $\be_3\rightarrow0$ (see the discussion in \cite{IvKarl13pra}). However, this value of the velocity is out of range of applicability of the formula since it contradicts the assumptions and approximations made in deriving \eqref{dP_phi1}. Recall that, in deriving the probability \eqref{dP_phi1}, we assumed that $\vf_s(\spp)$ is concentrated at negative $p^3$ and discarded the contributions where $\vf_s(\spp)$ is taken with positive $p^3$. For the wave packets with Gaussian profile along $p^3$, we implied thereby that $p_3\gg\s_3$, where $\s_3^2$ is the dispersion of $p^3$ in the wave packet.

The formula \eqref{dP_phi} allows us to estimate the magnitude of the dispersion of $\spp$ near the point \eqref{norm_fall} when the contribution \eqref{dP_phi1} still dominates. In the nonrelativistic limit, it follows from \eqref{dP_phi} and \eqref{nr_lim} that
\begin{equation}\label{estim_nonrel}
    \De\theta^2\ll\Big(\frac{k_0}{p_3\be_3}\Big)^2,
\end{equation}
where $\De\theta^2$ is the dispersion of angles of momenta to the vector \eqref{norm_fall} in the wave packet. In the ultrarelativistic case, the approximate formulas \eqref{ur_lim} imply
\begin{equation}\label{estim_ultra}
    \ga^2\De\theta^2\ll\frac{k_0^2}{p^2_3}(1+n_\perp^2\ga^2),\quad n_\perp\ll1;\qquad \De\theta^2\ll\frac{k_0^2}{p^2_3},\quad n_\perp\sim1/2.
\end{equation}
For example, for a nonrelativistic particle with the kinetic energy of order $1$ keV and the energy of a detected photon of order $1$ eV, the estimate for the dispersion becomes $\De\theta^2\ll10^{-6}$.

Now we sum the expression \eqref{dP_fin} over the polarizations of an escaping photon in the case of a normal falling of the wave packet onto the mirror \eqref{norm_fall}. Employing the formulas \eqref{polar_sum1}, \eqref{polar_sum2} for summation over the photon polarizations, we arrive at
\begin{equation}\label{summed_pol_gen}
    dP(\spk;\vf)=\int\frac{d\spp c(\spp)}{|p_3'p_3|}\Big[(e^2-q^2\mu_a^2)n_\perp^2\Big(\frac{2p_3q_3}{q^2}+1\Big)^2 -(e+2m\mu_a)^2\frac{2k_3^2}{q^2}\Big] \frac{d\spk}{16\pi^3k_0}.
\end{equation}
This expression is independent of the effective spin of an incident particle. Keeping only the leading terms with respect to the quantum recoil parameter \eqref{quant_rec}, we have
\begin{equation}\label{summed_pol}
    dP(\spk;\vf)\approx \int \frac{d\spp c(\spp)}{n_3^2+n_\perp^2\ga^2} \Big[ 2n_\perp^2\ga^2 \Big(\frac{e^2\be_3^2\ga^2}{k_0^2(n_3^2+n_\perp^2\ga^2)} +\mu_a^2\Big) +(e+2m\mu_a)^2 \frac{n_3^2}{m^2} \Big]   \frac{d\spk}{8\pi^3k_0},
\end{equation}
where $\ga=\ga(\spp)$ and $\be_3=\be_3(\spp)$.

The different contributions to the radiation probabilities \eqref{dP_trans0}, \eqref{summed_pol} can be classified by restoring the Planck constant in the corresponding intensities of radiation $k_0dP$. Supposing that
\begin{equation}\label{hbar_orders}
    e\sim\hbar^{-1/2},\qquad\mu_a \sim\hbar^{1/2},\qquad\spk\sim\hbar,\qquad k_0\sim\hbar,
\end{equation}
and the other quantities in \eqref{dP_trans0}, \eqref{summed_pol} are of zero order in $\hbar$, we see that the contribution \eqref{dP_trans0} is purely quantum and contains the terms of the orders $\hbar^2$, $\hbar^3$, and $\hbar^4$. The first term in the parentheses in \eqref{summed_pol} is classical and proportional to $\hbar^0$, the second term in these brackets is of the order $\hbar^4$, and the last term in the square brackets in \eqref{summed_pol} contains the contributions of the orders $\hbar^2$, $\hbar^3$, and $\hbar^4$. Comparing \eqref{summed_pol} with \eqref{dP_trans0}, we see that the contribution \eqref{dP_trans0} is solely due to quantum recoil and is of the order $(k_0/p_3\be_3)^2$ as compared with the leading classical contribution to the probability \eqref{summed_pol}.

\section{Examples of the wave packets}\label{WavePack}

\subsection{Coherent superposition of $N$ Gaussians}

The profile of the wave packet can be taken as a superposition of $N$ Gaussians:
\begin{equation}\label{Schr_cats}
    c(\spp)=|\vf(\spp)|^2,\qquad\vf(\spp)=k e^{-g^{ij}(p_i-p_i^0)(p_i-p_j^0)/4}\sum_{l=1}^Nk_le^{-i\spp\mathbf{b}_l},
\end{equation}
where the matrix $g_{ij}$ is inverse to $g^{ij}$ and specifies the dispersions of momenta with respect to the average value $\spp_0$. The normalization constant is
\begin{equation}
    k^{-2}=(2\pi)^{3/2}\sqrt{\det g_{ij}}\sum_{k,l=1}^N k_kk^*_le^{-i\spp_0 \mathbf{b}_{kl}}e^{-g_{ij}b^i_{kl}b^j_{kl}/2},\qquad \mathbf{b}_{kl}:=\mathbf{b}_{k}-\mathbf{b}_{l}.
\end{equation}
The Gaussians in \eqref{Schr_cats} are transformed to each other by a parallel transport in the coordinate space along the vectors $\mathbf{b}_k$. The constants $k_i$ define the weights and the relative phases of these Gaussians. The spin polarization of the state is specified by some density matrix $\rho_{s,\bar{s}}(\spp,\spp)$.

One can achieve a coherent amplification of radiation created by a wave packet, analogous to the coherent amplification of radiation from many charged particles, by means of a special preparation of the wave packet with peculiar structure factor
\begin{equation}
    S(\spp)=\sum_{l,n=1}^Nk_lk_n^*e^{-i\spp\mathbf{b}_{ln}}.
\end{equation}
If all the $k_l$ are the same, the coherent amplification appears in the case when, for example, $\mathbf{b}_l$ form a periodic lattice with periods much larger than the square root of the eigenvalues of $g^{ij}$. Then $c(\spp)$ is peaked near the points $\spp_n$ of the reciprocal lattice and is modulated by the slowly varying Gaussian, the square root of which stands before the sum sign in \eqref{Schr_cats}. The harmonics in the energy of radiated photons are absent in this case in contrast to the radiation from many particles \cite{MarcuseI}. If one takes into account only the contributions of the maxima of $c(\spp)$ in \eqref{dP_fin}, then the radiation probability \eqref{dP_fin} has the form of the sum of probabilities corresponding to each maximum of $c(\spp)$, i.e., its form resembles the formula for an incoherent radiation from particles with different momenta $\spp_n$ \cite{PMHK08}.

\subsection{Spin polarized states}\label{SpinPolSt}

As we saw in Sec. \ref{RadWavPack}, the inclusive probability of transition radiation \eqref{dP_fin} depends on the effective spin of the initial Dirac particle state. Let us consider the effective spin of the state in more detail. The nontrivial example of the spin polarized wave packet of a Dirac particle is the state
\begin{equation}\label{Schr_cats_spin}
    \vf_s(\spp)=\frac{\vf_0(\spp)}{\sqrt{2}}e^{-is\psi/2},\qquad \psi:=\spp \mathbf{b}+\vartheta,
\end{equation}
where $s=\pm1$, the vector $\mathbf{b}$ specifies the displacement of the wave functions $\vf_0(\spp)$,
\begin{equation}
    \int d\spp|\vf_0(\spp)|^2=1,
\end{equation}
with opposite projections of the spin, and $\vartheta$ is the additional relative phase of these wave functions. This phase may appear, for example, due to different orbital angular momenta of the wave function components of the state \eqref{Schr_cats_spin} or it may arise in preparation of this state.

Then
\begin{equation}
    c(\spp)=|\vf_0(\spp)|^2,\qquad\rho_{s\bar{s}}(\spp,\spp)=\frac{1}{2}e^{-i(s-\bar{s})\psi/2}.
\end{equation}
The spin density matrix reads (see \eqref{mode_fun_Dir})
\begin{equation}\label{dens_matr_spin}
    \rho_{ab}(\spp,\spp)=\sum_{s,\bar{s}}\rho_{s\bar{s}}(\spp,\spp)\chi_{as}\chi^*_{b\bar{s}},\qquad a,b=\overline{1,2},
\end{equation}
where
\begin{equation}\label{spinors_bas}
    \bs{\s}\bs{\tau}\chi_s=s\chi_s.
\end{equation}
The unit vector
\begin{equation}
    \bs{\tau}=(\sin\theta\cos\phi,\sin\theta\sin\phi,\cos\theta)
\end{equation}
specifies the direction of the spin projection in the definition of the state \eqref{mode_fun_Dir}, \eqref{Schr_cats_spin}. The Hermitian normalized spin density matrix \eqref{dens_matr_spin} can be cast into the form
\begin{equation}\label{spin_dens_matr}
    \rho(\spp,\spp)=(1+\bs{\s}\bs{\zeta}_0(\spp))/2,
\end{equation}
where $\bs{\zeta}_0$ is the effective spin vector entering the right-hand side of \eqref{mix_state}. Using the explicit representation of the basis spinors \eqref{spinors_bas},
\begin{equation}
    \chi_+=
    \left[
      \begin{array}{c}
        \cos(\theta/2) \\
        e^{i\phi}\sin(\theta/2) \\
      \end{array}
    \right],\qquad
    \chi_-=
    \left[
      \begin{array}{c}
        -e^{-i\phi}\sin(\theta/2) \\
        \cos(\theta/2) \\
      \end{array}
    \right],
\end{equation}
it is not difficult to verify that
\begin{equation}
    \bs{\zeta}_0=\big(\cos\theta\cos\phi\cos(\psi-\phi)-\sin\phi\sin(\psi-\phi), \cos\theta\sin\phi\cos(\psi-\phi)+\cos\phi\sin(\psi-\phi), -\sin\theta\cos(\psi-\phi)\big).
\end{equation}
This vector obeys the relations
\begin{equation}
    \bs{\zeta}^2_0=1,\qquad \bs{\zeta}_0\bs{\tau}=0,\qquad\frac{d}{d\psi}\bs{\zeta}_0=[\bs{\tau},\bs{\zeta}_0].
\end{equation}
In particular, for $\theta=0$ we have
\begin{equation}
    \bs{\zeta}_0\big|_{\theta=0}=(\cos\psi,\sin\psi,0).
\end{equation}
In other words, the relative phase shift of the wave function components of the state \eqref{Schr_cats_spin} gives rise to rotation of the effective spin vector $\bs{\zeta}_0$ in the plane orthogonal to the vector $\bs{\tau}$. In particular, such a rotation happens when the vector $\mathbf{b}$ characterizing a relative displacement of the wave function components in \eqref{Schr_cats_spin} changes.

The above considerations are easily generalized to the state
\begin{equation}\label{Schr_cats_spin1}
    \vf_s(\spp)=\frac{\vf_0(\spp)}{\sqrt{2\cosh\kappa}}e^{s(\kappa-i\psi)/2},\qquad\kappa,\,\psi\in \mathbb{R},
\end{equation}
where $\kappa=\kappa(\spp)$ and $\psi=\psi(\spp)$. In fact, the state \eqref{Schr_cats_spin1} is the general form of a pure state of a Dirac particle. Then
\begin{equation}
    c(\spp)=|\vf_0(\spp)|^2,\qquad\rho_{s\bar{s}}=\frac{1}{2\cosh\kappa}e^{\kappa(s+\bar{s})/2-i\psi(s-\bar{s})/2}.
\end{equation}
The spin density matrix takes the form \eqref{spin_dens_matr} with
\begin{equation}\label{zeta_eff}
    \bs{\zeta}=\bs{\tau}\tanh\kappa+\bs{\zeta}_0/\cosh\kappa,\qquad\bs{\zeta}^2=1.
\end{equation}
In changing the phase $\psi$, the effective spin vector $\bs\zeta$ precesses around the vector $\bs{\tau}$. The free evolution of the state \eqref{Schr_cats_spin1} leaves this vector intact.

Thus we see that the transition radiation produced by such a wave packet is the same as from the particle in the quantum state with the density matrix whose diagonal in the momentum space has the form
\begin{equation}\label{diag_dens_matr}
    |\vf_0(\spp)|^2(1+\bs{\s}\bs{\zeta}(\spp))/2.
\end{equation}
The measurement of the spin dependent contributions to the probability \eqref{dP_fin} allows one to find a relative phase of the wave function components and the parameter $\kappa$ in the state \eqref{Schr_cats_spin1}. Such a situation occurs for any process
\begin{equation}
    \al+\be\rightarrow\ga',
\end{equation}
where the momentum conservation law holds and the states $\be$ and $\ga'$ possess definite momenta. Here $\al$ characterizes the initial state of a Dirac particle, $\be$ is the initial state of other particles, and $\ga'$ is some final state. It follows from the momentum conservation law that the corresponding transition probability depends only on the diagonal of the density matrix of the Dirac particle in the momentum space. This is valid for inclusive probabilities as well where the transition probability is summed over a subset of final states. Of course, it is impossible to create or to detect a particle with a given momentum because, due to the uncertainty relation, the installation of an infinite volume is needed for that. Nevertheless, in order for the above mentioned properties to hold, one can create the state $\be$ and record the state $\ga'$ that are narrow wave packets in the momentum space and such that their sizes in the coordinate space are larger than the size of the wave packet characterizing the state $\al$. In the reaction zone, the wave packet of the state $\al$ should be completely covered with the wave packets of the states $\be$ and $\ga'$.

In this respect, the mention should be made that the above relations for the effective spin polarization of a Dirac particle take place for photon states as well. Namely, the normalized momentum space diagonal of the photon density matrix (the spin density matrix) can be cast into the form
\begin{equation}\label{spin_dens_matr_phot}
    \sum_{\la,\la'}\rho_{\la\la'}(\spp,\spp)f^{(\la)}_i(\spp)f^{(\la')}_j(\spp),\qquad\rho=(1+\bs{\s}\bs{\zeta})/2,
\end{equation}
where $\mathbf{f}^{(\la)}(\spp)\in \mathbb{R}$ are the linear polarization vectors, $\{\mathbf{f}^{(1)}(\spp),\mathbf{f}^{(2)}(\spp),\spp\}$ constituting a right-handed triple, and $\bs{\zeta}$ is the Stokes vector, $|\bs{\zeta}|\leqslant1$. Let the initial state of the photon be represented in the form \eqref{Schr_cats_spin1}, where $s$ specifies the sign of the Stokes vector $\bs{\tau}$ for the basis components of the photon state. This state corresponds to the spin density matrix
\begin{equation}
    \rho_{\la\la'}(\spp,\spp)=\sum_{s,\bar{s}}\vf_s(\spp)\vf^*_{\bar{s}}(\spp)\chi_{\la s}\chi^*_{\la' \bar{s}}/c(\spp),\qquad c(\spp)=|\vf_0(\spp)|^2,
\end{equation}
where $\chi_s$ are defined in \eqref{spinors_bas}. Developing the considerations analogous to those given for a Dirac particle, we further infer that the spin density matrix takes the form \eqref{spin_dens_matr_phot} with the effective Stokes vector \eqref{zeta_eff} and the probability of the process complying with the above mentioned restrictions is determined by the momentum space diagonal of the density matrix \eqref{diag_dens_matr}. In a certain sense, one may say that a photon and a Dirac particle behave as a qubit in such processes.

For example, if $\bs{\tau}=(0,0,1)$ and $\kappa=0$, then the effective Stokes vector $\bs\zeta$ is orthogonal to $\bs\tau$ and, depending on the relative phase $\psi$, describes the circular, $\bs{\zeta}=(0,\pm1,0)$, elliptic, or linear, $\bs{\zeta}=(\pm1,0,0)$, polarizations, despite the fact that the wave function components $\vf_s(\spp)$ possessing the linear polarization along $\mathbf{f}^{((3-s)/2)}$ can be shifted with respect to each other by a large distance. For $\bs{\tau}=(0,1,0)$ and $\kappa=0$, the wave function components $\vf_s(\spp)$ are circularly polarized, whereas the effective Stokes vector $\bs\zeta$ is orthogonal to $\bs\tau$ and corresponds to the linear polarization at an angle of $\psi/2$ to the axis $\mathbf{f}^{(1)}$.

\subsection{Twisted particles}\label{TwisPart}

Let the state of a Dirac particle be specified by the wave packet of the form \cite{KarlZhev19,BliokhVErev,LBThY,Karl18}
\begin{equation}\label{twisted_Dir}
\begin{gathered}
    \vf_s(\spp)=\de_{s\s}\vf_0(\spp),\qquad \vf_0(\spp)=kp_\perp^{|l|}\exp\Big(-\frac{(p_3-p)^2}{4\s_3^2} -\frac{\spp_\perp^2}{4\s^2_\perp}+il\psi \Big),\\
    \psi:=\arg(p^1+ip^2),\qquad k^{-2}=(2\pi)^{3/2}\s_3\s_\perp^2(2\s_\perp^2)^{|l|}|l|!,\qquad l\in \mathbb{Z},
\end{gathered}
\end{equation}
where $\s=\pm1$ characterizes the spin projection of the state onto the $z$ axis. The state \eqref{twisted_Dir} is the eigenvector of the projection of the total angular momentum operator on the $z$ axis with the eigenvalue $l+\s/2$. The case $l=0$ describes a cylindrically symmetric Gaussian wave packet. Then
\begin{equation}
    c(\spp)=k^2p_\perp^{2|l|}\exp\Big(-\frac{(p_3-p)^2}{2\s_3^2} -\frac{\spp_\perp^2}{2\s^2_\perp}\Big).
\end{equation}
As is seen from the general formula \eqref{dP_fin} for the probability of transition radiation from the mirror, this probability does not depend on the phase of the wave function \eqref{twisted_Dir}.

The classical current \eqref{class_curr} corresponding to the state \eqref{twisted_Dir} without the phase $il\psi$ has zero projection of the intrinsic magnetic moment on the $z$ axis by symmetry reasons. Nevertheless, the nonzero angular momentum of the state affects the properties of the transition radiation \eqref{dP_fin} by means of the factor $p_\perp^{|l|}$ in the wave function. Let us find this correction. Setting $\mu_a=0$, disregarding the terms that are small with respect to the quantum recoil parameter \eqref{quant_rec}, and using formulas \eqref{contractions}, we deduce the probability to detect a linearly polarized photon
\begin{equation}\label{dP_twist_e}
    dP(\la,\spk;\vf)\approx e^2\int\frac{d\spp c(\spp)}{p_3^2(q^2)^2}|k_3(\spp \mathbf{f})+(q_3-k_3)p_3f_3|^2\frac{d\spk}{4\pi^3k_0},
\end{equation}
where the approximate expression for $q^2$ is presented in \eqref{small_reco}. This probability does not depend on the spin of the Dirac particle and, in particular, is valid for the mixed state where $\s=\pm1$ are realized with equal probability. In the last case, formula \eqref{dP_twist_e} holds for a radiated photon with arbitrary polarization.

If $(|l|+1)\s^2_\perp\ll m^2$ and $\s_3\ll|p|$, then the integral \eqref{dP_twist_e} is readily evaluated. Let
\begin{equation}
    \mathbf{n}=(\sin\theta,0,\cos\theta),\qquad \mathbf{f}^{(1)}=(\cos\theta,0,-\sin\theta).
\end{equation}
Then
\begin{equation}\label{dP1}
\begin{split}
    dP(1,\spk;\vf)\approx&\, \frac{e^2p^2}{(m^2+n_\perp^2p^2)^2}\Big\{n_\perp^2(m^2+p^2)+\\
    &+\frac{(|l|+1)\s_\perp^2}{(m^2+n_\perp^2p^2)^2} \big[m^4(1-10n_\perp^2+10n_\perp^4) -2m^2p^2n_\perp^2(5-7n_\perp^2+n_\perp^4) +p^4 n_\perp^4 \big] \Big\} \frac{d\spk}{4\pi^3k_0^3},
\end{split}
\end{equation}
where the leading correction with respect to $(|l|+1)\s^2_\perp$ is retained. Notice that the non-paraxial corrections proportional to $(|l|+1)\s^2_\perp/m^2$ are typical for the processes with twisted particles (see for details \cite{Karl18}). These corrections are enhanced for the states with a large projection of the orbital angular momentum. If
\begin{equation}
    \mathbf{f}^{(2)}=(0,1,0),
\end{equation}
we have in the leading order
\begin{equation}\label{dP2}
    dP(2,\spk;\vf)\approx e^2\frac{p^2 (|l|+1)\s_\perp^2}{(m^2+n_\perp^2p^2)^2} \frac{n_3^2d\spk}{4\pi^3k_0^3}.
\end{equation}
We see that the non-paraxial correction is the leading contribution in this case. The leading contribution to \eqref{dP1} is the standard expression for the probability of transition radiation from the mirror \cite{Ginzburg}.

The expressions \eqref{dP1} and \eqref{dP2} make sense only when the obtained corrections with respect to $\s_\perp$ are not overlapped by the corrections due to quantum recoil. The estimate assuring that the contribution \eqref{dP2} dominates over the corrections due to quantum recoil follows from the comparison of \eqref{dP2} with \eqref{dP_trans}. The contribution \eqref{dP2} prevails when the estimates opposite to \eqref{estim_nonrel}, \eqref{estim_ultra} are fulfilled. In the case at hand, this estimate can be cast into the form
\begin{equation}\label{estim_twist}
    (|l|+1)\s^2_\perp/m^2\gg k_0^2 (n_3^2+n_\perp^2\ga^2)/p^2.
\end{equation}
For example, for the electron wave packet with transverse size of the order $10$ nm, the projection of the orbital momentum $|l|\sim10$, and the photon energy $1$ eV, the electron kinetic energy should be larger than $600$ keV for the estimate \eqref{estim_twist} to be satisfied and the term \eqref{dP2} to dominate. Notice that, in contrast to the leading contribution to \eqref{dP1}, the obtained corrections to $dP/d\spk$ do not vanish for $n_\perp\rightarrow0$. This property can also be employed for a possible experimental verification of the presence of these corrections.

%when
%\begin{equation}\label{nonparax_recoil}
%    (|l|+1)\s^2_\perp/m^2\gg k_0/(p_3\be_3),
%\end{equation}
%i.e.,

% we have
%\begin{equation}
%    (|l|+1)\s^2_\perp/m^2\approx 1.5\times 10^{-8}.
%\end{equation}
%Then the estimate \eqref{nonparax_recoil} is satisfied for the photon energy $1$ eV and the electron energy $p_0\gtrsim 67$ MeV.

Let us find the transition radiation from the mirror produced by a twisted neutron \cite{Clark15,CapJachVin18,Sarena18,Sarena19}. We suppose that the state of the neutron has the form \eqref{twisted_Dir}. Putting $e=0$ and neglecting the quantum recoil, we find the probability to detect a linearly polarized photon
\begin{equation}\label{dPn}
    dP(\la,\spk;\vf)\approx -\mu_a^2\int\frac{d\spp c(\spp)}{p_3^2q^2}[m^2k_3^2 + |k_3(\spp \mathbf{f})+(q_3-k_3)p_3f_3|^2]\frac{d\spk}{4\pi^3k_0}.
\end{equation}
If the neutron state possesses a natural spin polarization, then this expression is valid for the photons with arbitrary polarizations. Using the same approximations as above in considering a charged twisted particle, we arrive at
\begin{equation}\label{dPn1}
    dP(1,\spk;\vf)=\mu_a^2\Big(1-\frac{(|l|+1)\s_\perp^2n_3^2}{m^2+n_\perp^2p^2} \Big) \frac{d\spk}{4\pi^3k_0},
\end{equation}
and
\begin{equation}\label{dPn2}
    dP(2,\spk;\vf)=\frac{\mu_a^2}{m^2+n_\perp^2p^2}\Big\{m^2-\frac{(|l|+1)\s_\perp^2}{(m^2+n_\perp^2p^2)^2}\big[m^4(1-3n_\perp^2) -m^2p^2n_\perp^2(4-n_\perp^2) -p^4n_\perp^4 \big] \Big\} \frac{n_3^2d\spk}{4\pi^3k_0}.
\end{equation}
Notice that the leading contribution to \eqref{dPn1} does not depend on the form of the wave packet provided $\s_\perp$ is sufficiently small. For example, this contribution has the same form for the state \eqref{twisted_Dir} with two Gaussian humps with respect to the variable $p_3$. It is also clear that the radiation described by this contribution is isotropic. As a rule, the correction to this term is small. Of course, these properties hold only in the applicability domain of the formula \eqref{dPn1}. The sum of the leading contributions to \eqref{dPn1} and \eqref{dPn2} coincides with \eqref{summed_pol} for $e=0$ and do not coincide with formula (9) of \cite{IvKarl13pra}.

Now we can assign the orders in $\hbar$ to the terms in \eqref{dP1}, \eqref{dP2}, \eqref{dPn1}, and \eqref{dPn2} using the rules \eqref{hbar_orders} and supposing that $\s_\perp\sim\hbar^{1/2}$. The last estimate follows from the assumption that $\s_\perp L_\perp\sim\hbar$, where $L_\perp\sim\hbar^{1/2}$ is some length characterizing the transverse size of the wave packet. In that case, one has a particle with the definite coordinate $\spx_\perp$ and momentum $\spp_\perp$ in the limit $\hbar\rightarrow0$ \cite{BBTq1,BBTq2}. Then the leading term in \eqref{dP1} is classical, the correction to it is of the order $\hbar$, and the contribution \eqref{dP2} to the radiation intensity is of the order $\hbar$. The first and second terms in the brackets in \eqref{dPn1}, \eqref{dPn2} are of the orders $\hbar^4$ and $\hbar^5$, respectively. Thus, in such a classification scheme, the corrections due to finite size of the wave packet are of the order $\hbar$ with respect to the leading contribution. If one assumes that $\s_\perp\sim 1$, i.e., $L_\perp\sim\hbar$ and tends to zero as $\hbar\rightarrow0$, then the finite size corrections to the radiation probability will be of the same order in $\hbar$ as the leading contributions. In the case $\s_\perp\sim \hbar$, the powers of $\hbar$ in these formulas change in an obvious manner.

In the ultrarelativistic limit, the expressions for the leading contributions to \eqref{dPn1} and \eqref{dPn2} turn into formulas (3.276) and (3.277) of \cite{Ginzburg} where one should take $\omega_{pe,1}^2\rightarrow\infty$, $\omega_{pe,2}^2\rightarrow0$, and $\theta_0=\pi/2$. We see that the anomalous magnetic moment behaves as a ``true'' magnetic moment \cite{BordTerBagSpL}. Notice that the general formulas for the ultrarelativistic limit of probability of transition radiation from neutral plane-wave Dirac particles with anomalous magnetic moment were considered in \cite{GrimNuef95,SakKur95}.

\section{Conclusion}

Let us summarize the results. We obtained the inclusive probability \eqref{dP_fin} to record a plane-wave photon created in transition radiation by a wave packet of one Dirac particle traversing an ideal conductor plate. The probability to detect a photon in such a process disregarding the photon polarization has a rather simple form \eqref{summed_pol_gen} in the case of normal incidence. It was supposed in deriving these formulas that the Dirac particle possesses the electric charge $e$ and the anomalous magnetic moment $\mu_a$. The calculations were performed in the leading order in the coupling constants $e$ and $\mu_a$. The diffraction of the Dirac particles on the atoms of the mirror plate and bremsstrahlung were neglected in describing this radiation for the given energies of incident particles and detected photons. The effects due to the transition layer between vacuum and the mirror plate were also disregarded.

Then we investigated several particular cases of the general formula \eqref{dP_fin}. We found that in contrast to the classical formula for transition radiation there are radiated photons possessing the linear polarization along the vector $[\spp,\spk]$, where $\spp$ is the Dirac particle momentum and $\spk$ is the photon momentum, i.e., possessing the linear polarization orthogonal to the reaction plane. This effect stems from both a quantum recoil and a finite size of the wave packet. As for Vavilov-Cherenkov radiation, the fact that the quantum recoil leads to production of photons with such a polarization was mentioned in \cite{IvSerZay16}. We obtained the probability \eqref{dP_phi1} to detect the plane-wave photons with such a polarization in the case of normal incidence of the wave packet of a Dirac particle onto the mirror plate. We also discussed the restrictions on the parameters of the wave packet when this contribution to radiation can be observed in experiments.

As the examples of wave packets, we considered the coherent superposition of Gaussians, the general spin polarized state, and the twisted states. In the case of $N$ Gaussians comprising a one-particle wave packet, the coherent amplification of transition probability could be achieved when these Gaussians constitute a regular lattice. It turned out in this case that the radiation probability is approximately an incoherent sum of radiation probabilities corresponding to particles with momenta $\spp_n$ of the reciprocal lattice. In contrast to the classical coherent radiation, there are not harmonics in the energy of radiated photons \cite{MarcuseI}. However, as it was discussed in Appendix \ref{ClassAppr}, the harmonics in energy reappear in the coherent radiation from $N$-particle wave packet consisting of such modulated one-particle wave packets. Besides, the harmonics in energy can be observed in the probability of the exclusive process where the escaping Dirac particle is measured in the $out$-state that is obtained by a free evolution from the $in$-state of this particle.

As far as the general spin polarized state is concerned, it behaves as a qubit in regard to the spin dependence of transition radiation. Namely, due to the fact that only the momentum space diagonal of the density matrix of the incident particle contributes to the transition radiation probability, one can introduce the effective spin vector of the initial state that takes the values on the Poincar\'{e} sphere for pure states. This effective spin vector is determined by the choice of the spin quantization axis of the wave packet and by the relative phase and the ratio of moduli of the respective spin components of this wave packet in the momentum representation. The effective spin vector can be rotated along the Poincar\'{e} sphere by changing these parameters, for example, by changing the distance between the wave packet components with opposite spin projections.

As for the twisted states of Dirac particles, we obtained the explicit expressions \eqref{dP1}, \eqref{dP2}, \eqref{dPn1}, \eqref{dPn2} for the probability to detect a plane-wave photon produced by one twisted particle traversing normally the conducting plate. The first nontrivial corrections in $\s_\perp^2$ were taken into account, where $\s_\perp^2$ is the dispersion of the transverse momentum components. The cases $\mu_a=0$ and $e=0$ were separately considered. As expected, the leading order contributions coincide with the known results \cite{Ginzburg} in the limit of a small quantum recoil. The corrections to this contribution are proportional to $(|l|+1)\s_\perp^2/m^2$ what is typical for processes with twisted particles \cite{Karl18}. We estimated the parameter space where these corrections can be observed experimentally for transition radiation from electrons. In the case of electrons, the non-paraxial correction to the probability of transition radiation of photons with the polarization vector orthogonal to the reaction plane is the leading contribution. This makes it easier to observe such a correction experimentally. We also revealed an interesting feature of the transition radiation produced by a neutron. In this radiation, the probability \eqref{dPn1} to detect a photon with polarization vector lying in the plane spanned by the vectors $\spp$ and $\spk$ is solely determined by the anomalous magnetic moment $\mu_a$ and does not depend on the observation angle and the energy of the incident neutron for sufficiently small $\s_\perp^2$. It would be interesting to check this property experimentally.

Inasmuch as transition radiation from twisted particles was already discussed in the literature \cite{IvKarl13prl,IvKarl13pra,KonPotPol,BliokhVErev} from the viewpoint of classical radiation theory applied to the Dirac current of a wave packet, in Appendix \ref{ClassAppr} we compared in detail the method developed in the present paper with the method used in \cite{IvKarl13prl,IvKarl13pra,KonPotPol}. We found that the inclusive probability \eqref{dP_fin} cannot be reproduced by the classical approach applied to radiation from a one-particle wave packet. Nevertheless, we described the experimental setup where the classical approach does give the correct result for transition radiation from a one-particle wave packet. Moreover, we showed that under certain mild assumptions the coherent radiation from $N$-particle wave packets is determined by the classical radiation amplitudes associated with the $N$ wave packets of particles constituting the $N$-particle state.

\paragraph{Acknowledgments.}

We appreciate D.V. Karlovets for useful comments and suggestions. The reported study was supported by the Russian Ministry of Education and Science, the contract N 0721-2020-0033.

\appendix
\section{Classical description of radiation from a wave packet}\label{ClassAppr}

In the papers \cite{IvKarl13prl,IvKarl13pra,KonPotPol}, transition radiation from the wave packet of a twisted electron was studied using a classical approach. Namely, the radiation from such a wave packet was modeled by means of the radiation from a point charged classical particle with large magnetic moment. The reasoning behind this approach is the assumption that the transition radiation is produced by the classical current,
\begin{equation}\label{class_curr}
    j^i(x)=e\bar{\vf}(x)\ga^i\vf(x),
\end{equation}
constructed by making use of the solutions $\vf(x)$ of the Dirac equation. If one substitutes $\vf(x)$ in the form of a twisted electron into \eqref{class_curr}, then the classical electromagnetic field (or the average of the quantum field operator) created by $j^i(x)$ at large distances from the electron is approximately the same as for a point charged magnetic moment. The small contributions of higher multipoles are also present \cite{KarlZhev19}. In this Appendix, we shall show how the expressions obtained in \cite{IvKarl13prl,IvKarl13pra,KonPotPol} can be interpreted in the framework of QED.

\subsection{One-particle wave packets}

Denote as
\begin{equation}
    \tilde{\phi}_{\al'}:=e^{-ip'_0t_2} \phi_{\al'}
\end{equation}
the electron wave function recorded by the detector at the instant of time $t_2$. Then $\phi_{\al'}$ gives the form of the electron wave packet at the instant of time $0$, as if the electron were freely propagating at $t\in[0,t_2]$. Such an assumption is valid in the leading order of the perturbation theory. The transition amplitude from the electron state $\tilde{\vf}$ to the electron state $\tilde{\phi}$ with radiation of one photon can easily be deduced from formula \eqref{ampl_phi} that gives
\begin{equation}\label{ampl_phi_phi}
\begin{split}
    A(\ga,\phi;\vf)=\,&\sum_{\al'}  \sqrt{\frac{(2\pi)^3}{V}}\tilde{\phi}^*_{\al'} A(\ga,\al';\vf)=\\
    =\,&iem e^{-iE_0(t_2-t_1)}e^{-ik_0t_2}\sum_{s,s'}\int\frac{d\spp d\spp'}{(2\pi)^3} \int_{t_1}^{t_2}dx   e^{ik_0x^0-i\spk_\perp \spx_\perp+i(p_\mu'-p_\mu)x^\mu}\times\\
    &\times \phi^*_{s'}(\spp')\frac{\mathbf{a}^*_\la(x_3)\bar{u}_{s'}(\spp')\bs{\ga} u_{s}(\spp)}{\sqrt{2Vk_0p_0p_0'}}\vf_s(\spp)=\\
    =\,&iee^{-iE_0(t_2-t_1)}e^{-ik_0t_2} \int_{t_1}^{t_2}dx \frac{\mathbf{a}^*_\la(x_3)\bar{\phi}(x)\bs{\ga} \vf(x)}{\sqrt{2Vk_0}} e^{ik_0x^0-i\spk_\perp \spx_\perp},
\end{split}
\end{equation}
where, on the last line, the coordinate representation of the Dirac wave functions was used,
\begin{equation}
    \vf(x):=\lan0|\hat{\psi}(x)|\vf\ran=\sum_{\al}  \sqrt{\frac{(2\pi)^3}{V}} \vf_{\al}  \lan0|\hat{\psi}(x)|\al\ran=\sum_s\int\frac{d\spp}{(2\pi)^{3/2}}\sqrt{\frac{m}{p_0}}u_s(\spp)\vf_s(\spp)e^{-ip_\mu x^\mu},
\end{equation}
and we assume that the anomalous magnetic moment $\mu_a=0$.

Suppose
\begin{equation}\label{joining}
    \phi_{\al}=\vf_{\al},
\end{equation}
that means that the electron is detected at $t=t_2$ in the state resulting from a free evolution of the state $\tilde{\vf}_\al$ prepared at $t=t_1$. Then taking the limits $t_1\rightarrow-\infty$, $t_2\rightarrow\infty$, and squaring the modulus of \eqref{ampl_phi_phi}, we see that the classical approach for description of the wave packet radiation gives the transition probability
\begin{equation}\label{dP_class}
    dP(\ga,\vf;\vf)=\Big| \int_{-\infty}^{\infty}dx \mathbf{a}^*_\la(x_3)\mathbf{j}(x) e^{ik_0x^0-i\spk_\perp \spx_\perp} \Big|^2 \frac{d\spk}{2(2\pi)^3k_0}.
\end{equation}
Thus, in the case of a one-particle wave packet, the ``classical'' formula for radiation probability possesses the following interpretation: It gives the probability to record a photon in the state $\ga$ in the remote future under the condition that, in the remote future, the electron is recorded in the state that results from a free evolution of the electron state prepared in the remote past. It is clear that
\begin{equation}
    dP(\ga,\vf;\vf)<dP(\ga;\vf).
\end{equation}
For such a setup of the experiment, the radiation from a one-particle wave packet possesses all the properties of the classical coherent radiation without corrections for a quantum recoil. In particular, if one prepares properly the state $\vf$, there exist the coherent harmonics in the energies of radiated photons. In a certain sense, one may say that recording the escaping particle in the state  $\tilde{\phi}_{\al'}$ that obeys the condition \eqref{joining} suppresses the quantum recoil produced by a photon radiated in this process. These considerations and the formula of the form \eqref{dP_class} also hold in the case when the spinors $u_\al$ are the solutions of the Dirac equation in a stationary external field and the mode functions of the electromagnetic field are the solutions of the Maxwell equations in a stationary dispersive medium. In that case, the free evolution of a wave packet ought to be understood as the evolution of solutions of the Dirac equation in a given field.

For certain energies of the observed photon and the incident particle, external fields, and profiles of the incident particle wave packet, the change of the state of this particle due to its interaction with photon can be neglected. Then the expression \eqref{dP_class} equals approximately to \eqref{dP_ini} (see, e.g., \cite{BaiKat1,BaKaStr,BaKaStrbook,AkhShul91,BBTq1,BBTq2,AkhShul,Bord.1,BKL4}). As for transition radiation from the mirror that we are considering in the present paper, the expression \eqref{dP_ini} is independent of the phase of the initial wave packet $\vf_s(\spp)$. Hence, \eqref{dP_class} must also be independent of this phase provided one intends to use \eqref{dP_class} as an approximation to \eqref{dP_ini} and not in the sense of the above given interpretation. However, this is not the case. For example, for a wave packet of a Dirac particle with nonzero projection of the orbital angular momentum \eqref{twisted_Dir}, the inclusive radiation probability is the same as for a wave packet without the phase $\psi$. The classical current \eqref{class_curr} corresponding to the wave packet with large projection of the orbital angular momentum possesses a large magnetic moment that leaves a typical imprint on radiation. On the other hand, the classical current \eqref{class_curr} corresponding to the wave packet with vanishing projection of the orbital angular momentum possesses a zero magnetic moment by symmetry reasons. Thus we see that formula \eqref{dP_class} cannot be employed as an approximation to the inclusive probability \eqref{dP_ini} in this case (see for details Sec. \ref{TwisPart}).

\subsection{$N$-particle wave packets}

Nevertheless, this is not the end of the story. We shall show now that under certain circumstances the classical formula \eqref{dP_class} reappears in the inclusive probability of radiation produced by $N$-particle wave packets. Let
\begin{equation}\label{Npart_state}
    c \vf_{\text{in}\,\al_1}^1\cdots \vf_{\text{in}\,\al_N}^N a^\dag_{\al_1}\cdots a^\dag_{\al_N}|0\ran
\end{equation}
be the initial state of the system at $t=t_1$. Here, for brevity, the normalization to unity and summation over repeated indices are understood. The constant $c$ is the normalization constant. In the Bargmann-Fock representation (see, e.g., \cite{BerezMSQ1.4,ippccb}), the state \eqref{Npart_state} is written as
\begin{equation}
    \Phi(\bar{a})=c(\vf^1_{\text{in}}\bar{a})\cdots (\vf^N_{\text{in}} \bar{a}),
\end{equation}
where $\bar{a}_\al$ are the anticommuting variables. The normalization condition becomes
\begin{equation}%\label{normal_cond}
    \int D\bar{a}Dae^{-(\bar{a}a)}\Phi(a)\Phi(\bar{a})=1,\qquad \int D\bar{a}Dae^{-(\bar{a}a)}=1,
\end{equation}
where the latter equality is the normalization condition for the functional integral and $\Phi(a)\equiv\bar{\Phi}(\bar{a})$ is a complex conjugate to $\Phi(\bar{a})$. For compliance with the notation of \cite{ippccb}, henceforth in this section the bar over an expression means a complex conjugation. Denote  as $A_{\al'\al}$ the matrix element \eqref{matrix_elem} in the leading order of the perturbation theory. Then, in this order of the perturbation theory, the inclusive probability to record a photon in the state $\ga$ produced in the process with the initial state \eqref{Npart_state} reads
\begin{equation}
    P(\ga)=\int D\bar{a}Da \Phi(a)\frac{\overleftarrow{\de}}{\de a_\be} \bar{A}_{\be'\be}a_{\be'}  e^{-(\bar{a}a)} \bar{a}_{\al'}  A_{\al'\al}\frac{\de}{\de\bar{a}_\al}\Phi(\bar{a}),
\end{equation}
where we have used the formulas (226) and (232) of \cite{ippccb} for the decomposition of unity and for the creation-annihilation operators in the Bargmann-Fock representation.

Taking into account that
\begin{equation}
    \int D\bar{a}Dae^{-(\bar{a}a)-(\bar{a}\eta)-(\bar{\eta}a)}=e^{(\bar{\eta}\eta)},
\end{equation}
where $\eta_\al$, $\bar{\eta}_\al$ are the anticommuting sources, we obtain
\begin{equation}
\begin{split}
    P(\ga)=\,&|c|^2\sum_{k,l=1}^N \Big(\bar{\vf}^N_{\text{in}}\frac{\de}{\de\bar{\eta}}\Big)\cdots \Big(\bar{\vf}^k_{\text{in}} A^\dag\frac{\de}{\de\bar{\eta}}\Big) \cdots \Big(\bar{\vf}^1_{\text{in}}\frac{\de}{\de\bar{\eta}}\Big) e^{(\bar{\eta}\eta)} \Big(\frac{\overleftarrow{\de}}{\de\eta}\vf^1_{\text{in}}\Big)\cdots  \Big(\frac{\overleftarrow{\de}}{\de\eta} A \vf^l_{\text{in}}\Big)\cdots \Big(\frac{\overleftarrow{\de}}{\de\eta}\vf^N_{\text{in}}\Big)\Big|_{\eta=\bar{\eta}=0}=\\
    =\,&|c|^2\sum_{k,l=1}^N \Big(\bar{\vf}^N_{\text{in}}\frac{\de}{\de\bar{\eta}}\Big)\cdots \Big(\bar{\vf}^k_{\text{in}} A^\dag\frac{\de}{\de\bar{\eta}}\Big) \cdots \Big(\bar{\vf}^1_{\text{in}}\frac{\de}{\de\bar{\eta}}\Big) \Big(\bar{\eta}\vf^1_{\text{in}}\Big)\cdots  \Big(\bar{\eta} A \vf^l_{\text{in}}\Big)\cdots \Big(\bar{\eta}\vf^N_{\text{in}}\Big).
\end{split}
\end{equation}
This expression can be rewritten in terms of the determinants,
\begin{equation}
\begin{split}
    |c|^{-2}=&\,\det(\bar{\vf}^i_{\text{in}}\vf^j_{\text{in}}),\\
    d_{lk}:=&\,\det
    \left[
      \begin{array}{ccccc}
        \bar{\vf}^1_{\text{in}}\vf^1_{\text{in}} & \cdots & \bar{\vf}^k_{\text{in}}A^\dag\vf^1_{\text{in}} & \cdots & \bar{\vf}^N_{\text{in}}\vf^1_{\text{in}} \\
        \vdots &   & \vdots &   & \vdots \\
        \bar{\vf}^1_{\text{in}}A\vf^l_{\text{in}} & \cdots & \bar{\vf}^k_{\text{in}}A^\dag A\vf^l_{\text{in}} & \cdots & \bar{\vf}^N_{\text{in}}A\vf^l_{\text{in}} \\
        \vdots &  & \vdots &  & \vdots \\
        \bar{\vf}^1_{\text{in}}\vf^N_{\text{in}} & \cdots & \bar{\vf}^k_{\text{in}}A^\dag\vf^N_{\text{in}} & \cdots & \bar{\vf}^N_{\text{in}}\vf^N_{\text{in}} \\
      \end{array}
    \right],
\end{split}
\end{equation}
as
\begin{equation}
    P(\ga)=|c|^2\sum_{k,l=1}^Nd_{lk}.
\end{equation}

In order to proceed, we assume that the wave packets $\vf^i$ are separated in space such that
\begin{equation}
    (\bar{\vf}^i_{\text{in}}\vf^j_{\text{in}})\approx\de_{ij},\qquad i,j=\overline{1,N}.
\end{equation}
Notice that these relations are conserved in time by a free evolution. Then
\begin{equation}
    |c|^2\approx1,
\end{equation}
and
\begin{equation}\label{P_ga}
    P(\ga)\approx\sum_{k=1}^N\bar{\vf}^k_{\text{in}}A^\dag A\vf^k_{\text{in}}+\sideset{}{'}\sum_{k,l=1}^N \big[(\bar{\vf}^k_{\text{in}} A^\dag\vf^k_{\text{in}}) (\bar{\vf}^l_{\text{in}} A\vf^l_{\text{in}}) -(\bar{\vf}^k_{\text{in}} A^\dag\vf^l_{\text{in}}) (\bar{\vf}^l_{\text{in}} A\vf^k_{\text{in}}) \big],
\end{equation}
where the prime at the sum sign means that the terms with $k=l$ are omitted. The second term in \eqref{P_ga} can be cast into the form
\begin{equation}\label{class_rad}
    \sideset{}{'}\sum_{k,l=1}^N (\bar{\vf}^k_{\text{in}} A^\dag\vf^k_{\text{in}}) (\bar{\vf}^l_{\text{in}} A\vf^l_{\text{in}})=\Big|\sum_{l=1}^N \bar{\vf}^l_{\text{in}} A\vf^l_{\text{in}}\Big|^2-\sum_{l=1}^N \big|\bar{\vf}^l_{\text{in}} A\vf^l_{\text{in}}\big|^2.
\end{equation}
Comparing the terms in \eqref{P_ga}, \eqref{class_rad} with \eqref{dP_ini}, \eqref{dP_class}, we infer the following interpretation of the contributions to \eqref{P_ga}: The first term in \eqref{P_ga} is the incoherent sum of the quantum contributions \eqref{dP_ini} to radiation from the wave packets $\vf^k$; the second term in \eqref{P_ga} is the classical contribution \eqref{dP_class} to radiation produced by the sum of the classical currents \eqref{class_curr} corresponding to the wave packets $\vf^k$, the incoherent contribution to the classical radiation should be excluded; the third term in \eqref{P_ga} is the exchange term.

Neglecting the exchange term, what is justified when, for example, $\vf^i$ are well separated for different $i$ at the instants of time when the radiation is forming, we arrive at
\begin{equation}
    P(\ga)\approx\sum_{k=1}^N\big(\bar{\vf}^k_{\text{in}}A^\dag A\vf^k_{\text{in}} -\big|\bar{\vf}^k_{\text{in}} A\vf^k_{\text{in}}\big|^2\big)+\Big|\sum_{k=1}^N \bar{\vf}^k_{\text{in}} A\vf^k_{\text{in}}\Big|^2.
\end{equation}
The standard theory of classical coherent radiation can be applied to describe the properties of the last term in this expression (cf. \cite{MarcuseII,SundMil90,PMHK08}).

Thus we see that, under the above assumptions, the classical contribution \eqref{dP_class} with the total current of $N$ wave packets determines the inclusive probability of coherent radiation from the $N$-particle wave packet. This term dominates provided the classical radiation amplitudes corresponding to the wave packets $\vf^k$ add up coherently and $N$ is large. For example, the transition radiation from a bunch train of twisted electrons with the wave functions obtained from each other by a parallel transport contains coherent harmonics. Then the radiation at these harmonics is approximately the coherent radiation from point charged magnetic moments with $\mu\approx l\mu_B$, where $l$ is a large projection of the orbital angular momentum of a one twisted electron \cite{IvKarl13prl,IvKarl13pra,KonPotPol,BliokhVErev}. Of course, due to unitarity, all the above formulas are also valid for absorption of photons by $N$-particle wave packets of fermions.

\section{Useful formulas}\label{UsefForm}

In the limit of a small quantum recoil, $k_0\ll p_3\be_3$, we have the relations
\begin{equation}\label{small_reco}
\begin{gathered}
    q_3\approx k_0(1-\bs{\be}_\perp\mathbf{n}_\perp)/\be_3,\qquad q_3-rk_3\approx k_0(1-\bs{\be}_\perp\mathbf{n}_\perp-r\be_3n_3)/\be_3,\\
    -q^2\approx k_0^2 \big[(1-\bs{\be}_\perp\mathbf{n}_\perp)^2-\be^2_3n^2_3 \big]/\be_3^2.
\end{gathered}
\end{equation}
In the nonrelativistic limit, $|\bs{\be}|\ll1$, we obtain
\begin{equation}\label{nr_lim}
    q_3\approx k_0/\be_3,\qquad q_3-rk_3\approx k_0/\be_3,\qquad -q^2\approx k_0^2/\be_3^2.
\end{equation}
In the ultrarelativistic limit, $\ga\gg1$, $\be_\perp\ll1$, $n_\perp\ll1$, we come to
\begin{equation}\label{ur_lim}
\begin{gathered}
    q_3\approx k_0,\qquad q_3-k_3\approx k_0(1+(\bs{\be}_\perp-\mathbf{n}_\perp)^2\ga^2)/(2\ga^2),\qquad q_3+k_3\approx 2k_0,\\
    -q^2\approx k^2_0[1+(\bs{\be}_\perp-\mathbf{n}_\perp)^2\ga^2]/\ga^2.
\end{gathered}
\end{equation}

Suppose that $f_3^{(\la)}(\spk)\in \mathbb{R}$. This can always be achieved by choosing appropriately the common phase of the polarization vector. Then
\begin{equation}\label{contractions}
\begin{split}
    \sum_r \frac{f_{ri}^{(\la)} }{q_3-rk_3}=\,&-\frac{2}{q^2}\big[k_3f_i^{(\la)} +(q_3-k_3)\de^3_if_3^{(\la)} \big],\\
    \sum_{r,r'} \frac{f_{ri}^{(\la)} f_{r'j}^{*(\la)}}{(q_3-rk_3)(q_3-r'k_3)}=\,&\frac{4}{(q^2)^2}\big[k_3^2f_i^{(\la)}f_j^{*(\la)} +k_3(q_3-k_3)(f_i^{(\la)}\de^3_j +\de^3_if_j^{*(\la)}) f_3+\\
    &+(q_3-k_3)^2\de^3_i\de^3_j (f_3)^2 \big],\\
    \sum_{r,r'}\frac{f_{ri}^{(\la)}p^{(i}p'^{j)} f_{r'j}^{*(\la)}}{(q_3-rk_3)(q_3-r'k_3)}=\,& \frac{8}{(q^2)^2} \big\{k_3^2|\spp \mathbf{f}|^2
    +k_3(p_3(q_3-k_3)-q^2/2)\spp(\mathbf{f}+\mathbf{f}^*)+\\
    &+[p_3^2(q_3-k_3)^2 -q^2p_3(q_3-k_3)]f_3^2  \big\},\\
    \sum_{r,r'}\frac{f_{ri}^{(\la)}\eta^{ij} f_{r'j}^{*(\la)}}{(q_3-rk_3)(q_3-r'k_3)}=\,&\frac{4}{(q^2)^2}\big[q^2 (f_3^{(\la)})^2-k_3^2 \big],\\
    \sum_{r,r'}\frac{f_{ri}^{(\la)}(p^i+p'^i)(p^j+p'^j)f_{r'j}^{*(\la)}}{(q_3-rk_3)(q_3-r'k_3)}=\,& \frac{16}{(q^2)^2}\big\{k_3^2|\spp \mathbf{f}|^2
    +k_3(p_3(q_3-k_3)-q^2/2)\spp(\mathbf{f}+\mathbf{f}^*)+\\
    &+(p_3(q_3-k_3)-q^2/2)^2f_3^2  \big\},\\
    \sum_{r,r'}\frac{f_{ri}^{(\la)}q^iq^j f_{r'j}^{*(\la)}}{(q_3-rk_3)(q_3-r'k_3)}=\,&4(f_3^{(\la)})^2.
\end{split}
\end{equation}
Besides, we need the formulas
\begin{equation}
\begin{gathered}
    \sum_r\frac{r}{q_3-rk_3}=-\frac{2k_3}{q^2},\qquad\sum_r\frac{1}{q_3-rk_3}=-\frac{2q_3}{q^2},\\
    \sum_{r,r'}\frac{rr'}{(q_3-rk_3)(q_3-r'k_3)}=\frac{4k_3^2}{(q^2)^2},\qquad \sum_{r,r'}\frac{1}{(q_3-rk_3)(q_3-r'k_3)}=\frac{4q_3^2}{(q^2)^2},
\end{gathered}
\end{equation}
and
\begin{equation}\label{polar_sum1}
\begin{gathered}
    \sum_\la (f_3^{(\la)})^2=n_\perp^2,\\
    \sum_{\la,r,r'}\frac{f_{ri}^{(\la)}\eta^{ij} f_{r'j}^{*(\la)}}{(q_3-rk_3)(q_3-r'k_3)}=\frac{4}{(q^2)^2}(q^2n_\perp^2-2k_3^2),\qquad
    \sum_{\la,r,r'}\frac{f_{ri}^{(\la)}q^iq^j f_{r'j}^{*(\la)}}{(q_3-rk_3)(q_3-r'k_3)}=4n_\perp^2.
\end{gathered}
\end{equation}
For $\spp=(0,0,p)$, we have
\begin{equation}\label{polar_sum2}
\begin{split}
    \sum_{\la,r,r'}\frac{f_{ri}^{(\la)}p^{(i}p'^{j)} f_{r'j}^{*(\la)}}{(q_3-rk_3)(q_3-r'k_3)}&= 8n_\perp^2\Big(\frac{p^2q_3^2}{(q^2)^2}-\frac{pq_3}{q^2}\Big)\approx 8n_\perp^2\frac{p^2q_3^2}{(q^2)^2},\\
    \sum_{\la,r,r'}\frac{f_{ri}^{(\la)}(p^i+p'^i)(p^j+p'^j)f_{r'j}^{*(\la)}}{(q_3-rk_3)(q_3-r'k_3)}&=4n_\perp^2\Big(\frac{4p^2q_3^2}{(q^2)^2}-\frac{4pq_3}{q^2}+1\Big)\approx 16n_\perp^2\frac{p^2q_3^2}{(q^2)^2},
\end{split}
\end{equation}
where, in the approximate equalities, the terms are disregarded that are small in the limit of a negligible quantum recoil.

%\newpage

\end{document}